\documentclass[mathleft,fleqn,%
]{an}
\usepackage{graphicx}
\usepackage[varg]{txfonts}
\newcommand{\msun}{$M_\odot$}

\begin{document}

\Pagespan{1}{}
\Yearpublication{2021}%
\Yearsubmission{2020}%
\Month{1}%
\Volume{342}%
\Issue{1}%
\DOI{asna.202100000}%

\title{Radius, rotational period and inclination of 
the Be stars in the Be/gamma-ray binaries MWC 148 and MWC 656\thanks{Data from TESS and Rozhen}}

\author{R. K. Zamanov\inst{1}\fnmsep\thanks{Corresponding author:
        {rkz@astro.bas.bg}}
\and  K. A. Stoyanov\inst{1}
\and  J. Mart\'{\i}\inst{2}
\and  V. D. Marchev\inst{1}
\and  Y. M. Nikolov\inst{1}
}
\titlerunning{The Be/gamma ray binaries MWC 148 and MWC 656}
\authorrunning{Zamanov et al.}
\institute{
Institute of Astronomy and National Astronomical Observatory, Bulgarian Academy of Sciences, \\ Tsarigradsko Shose 72, 
BG-1784 Sofia, Bulgaria
\and
Departamento de F\'isica, Escuela Polit\'ecnica Superior de Ja\'en, Universidad de Ja\'en, 
Campus Las Lagunillas,  A3, \\ 23071, Ja\'en, Spain 
}

\received{2020 November 3}
\accepted{2020 February 3}
\publonline{...}


\abstract{%
  Using TESS photometry and Rozhen spectra  of the Be/$\gamma$-ray binaries  MWC~148 and MWC~656,   
  we estimate  
  the projected rotational velocity ($ {v} \sin i$), 
  the  rotational period (P$_{\rm rot}$),  radius (R$_{\rm 1}$), and inclination ($i$) of the mass donor.
  For  MWC~148 we derive P$_{\rm rot} = 1.10 \pm 0.03$~d, R$_{\rm 1}= 9.2 \pm 0.5$~R$_\odot$, $i = 40^\circ \pm 2^\circ$,
			 and $ {v} \sin i =272 \pm 5$~km~s$^{-1}$. 
  For  MWC~656 we obtain P$_{\rm rot} = 1.12 \pm 0.03$~d, R$_{\rm 1}= 8.8 \pm 0.5$~R$_\odot$, $i = 52^\circ \pm 3^\circ$, 
			 and $ {v} \sin i =313 \pm 3$~km~s$^{-1}$.
  For  MWC~656 we also find that the rotation of the mass donor is coplanar with the orbital plane.  }
\maketitle

\section{Introduction}

The $\gamma$-ray binaries are a recently established and rare subclass of the high-mass X-ray binaries 
with most of their luminosity output being radiated 
 above 1 MeV (Dubus 2013; Chernyakova \& Malyshev 2020).
They are composed of an OBe donor star and a neutron star or a black hole (Mirabel 2012). 
The mechanism responsible for
the high-energy emission in these systems is still a subject of debate.
The $\gamma$-rays could be produced either by accretion-driven jets, or 
by the rotation-powered strong pulsar winds interacting with the nearby medium 
(Dubus 2006; Romero et al. 2007; Massi \& Jaron 2013),
and/or by a neutron star in the propeller regime (Wang \& Robertson 1985).

So far, seven systems have been confirmed as 
$\gamma$-ray binaries - PSR~B1259-63 (Aharonian et al. 2005a), LS~5039 (Aharonian et al. 2005b), LS~I~+61~303 (Albert et al. 2006),
HESS~J0632+057 (Aharonian et al. 2007), 1FGL~J1018.6-5856 (Corbet et al. 2011), PSR~J2032+4127 (Lyne et al. 2015) and LMC-P3 (Corbet et al. 2016). 
The nature of the compact object is already known in 
PSR~B1259-63 and PSR~J2032+4127, where radio observations confirm that
they contain neutron stars (Johnston et al. 1992; Abdo et al. 2009). 
Based on the donor star and the presence of a circumstellar disc, two subgroups of $\gamma$-ray binaries 
are proposed.  
The first subgroup harbours an O-type donor star and shows a single-peak profile in their $\gamma$-ray light-curve. 
The second subgroup contains an OBe star and shows several peaks in
their light-curves, occasionally correlated with the times 
when the compact object crosses 
and in some cases truncates the circumstellar disc of the donor star (Paredes \& Bordas 2019).

MWC~148 (HD 259440)  was identified as the optical counterpart of the variable TeV source HESS~J0632+057 (Aharonian et al. 2007) and also 
detected in the GeV domain (Li et al. 2017). 
The system consists of a B0Vpe star (Casares et al. 2012) and a compact object, 
with an orbital period P$_{\rm orb}$ = 315$^{+6}_{-4}$~d (Aliu et al. 2014). 

MWC~656 (HD 215227) is the suspected optical counterpart of the $\gamma$-ray source AGL~J2241+4454 
detected by the $AGILE$  satellite above 100 MeV  
(Lucarelli et al. 2010; Williams et al. 2010). 
 It is the first discovered binary containing a black hole as a companion of a Be star
(Casares et al. 2014). 
MWC~656 was only occasionally detected at GeV energies 
(Aleksi{\'c} et al. 2015). The black hole nature of the compact 
object  renders it  similar to the typical $\gamma$-ray binaries.
The orbital period of the system, obtained by optical photometry 
and later confirmed by radial velocity measurements, is P$_{\rm orb} = 60.37 \pm 0.04$~d 
(Williams et al. 2010; Casares et al. 2014). 

Here, using  photometric and spectral observations, we estimate  
the rotational period, the radius and the inclination of the mass donors
in  the Be/$\gamma$-ray binaries  MWC~148 and MWC~656. 

\begin{table*}
\centering
\caption{Spectroscopic observations of MWC~148 and MWC~656. In the table are given the date 
of observations,  UT start, exposure time in seconds, signal-to-noise ratio at about 6600 \AA, 
EW(H$\alpha$), FWHM($H\alpha$), EW(FeII5316), and FWZI(FeII5316).  
}  
\label{tab.j}
\begin{tabular}{ccrrccccccc}
 \hline
Object   &   Date-obs         &  exp-time & S/N  &  EW(H$\alpha$) & FWHM(H$\alpha$) & EW(FeII5316) & FWZI(FeII)    & \\
         & yyyy-mm-dd:hh:mm   &  [sec]    &	 &   [\AA]	  & [km~s$^{-1}$]   &  [\AA]	   & [km~s$^{-1}$] & \\
\\
 \hline 
MWC~148   &  2019-02-20T18:45 &	 2400 & 75   & -46.9 &  411  &  -0.59 & 672  & \\  
MWC~148   &  2019-12-06T01:45 &  2400 & 95   & -46.4 &  405  &  -0.60 & 711  & \\  
MWC~148   &  2020-01-16T21:16 &  3600 & 70   & -44.8 &  418  &  -0.59 & 683  & \\  %
MWC~148   &  2020-09-06T02:01 &  1800 & 60   & -43.7 &  412  &  -0.57 & 709  & \\  
& & & & & & & & \\	
MWC~656  &  2019-08-22T20:10  &  3600 &  90  & -20.9 &  493  &  -0.48 & 705 & \\   
MWC~656  &  2019-12-05T17:48  &  2400 & 100  & -21.6 &  491  &  -0.54 & 722 & \\   
MWC~656  &  2020-09-05T21:50  &  3600 &  64  & -21.7 &  490  &  -0.60 & 700 & \\
& & & & & & & & \\
 \hline  						             
 \end{tabular}  
 \label{tab.sp}						      
\end{table*}							      
 \begin{figure*}    
   \vspace{7.3cm}     
   \includegraphics{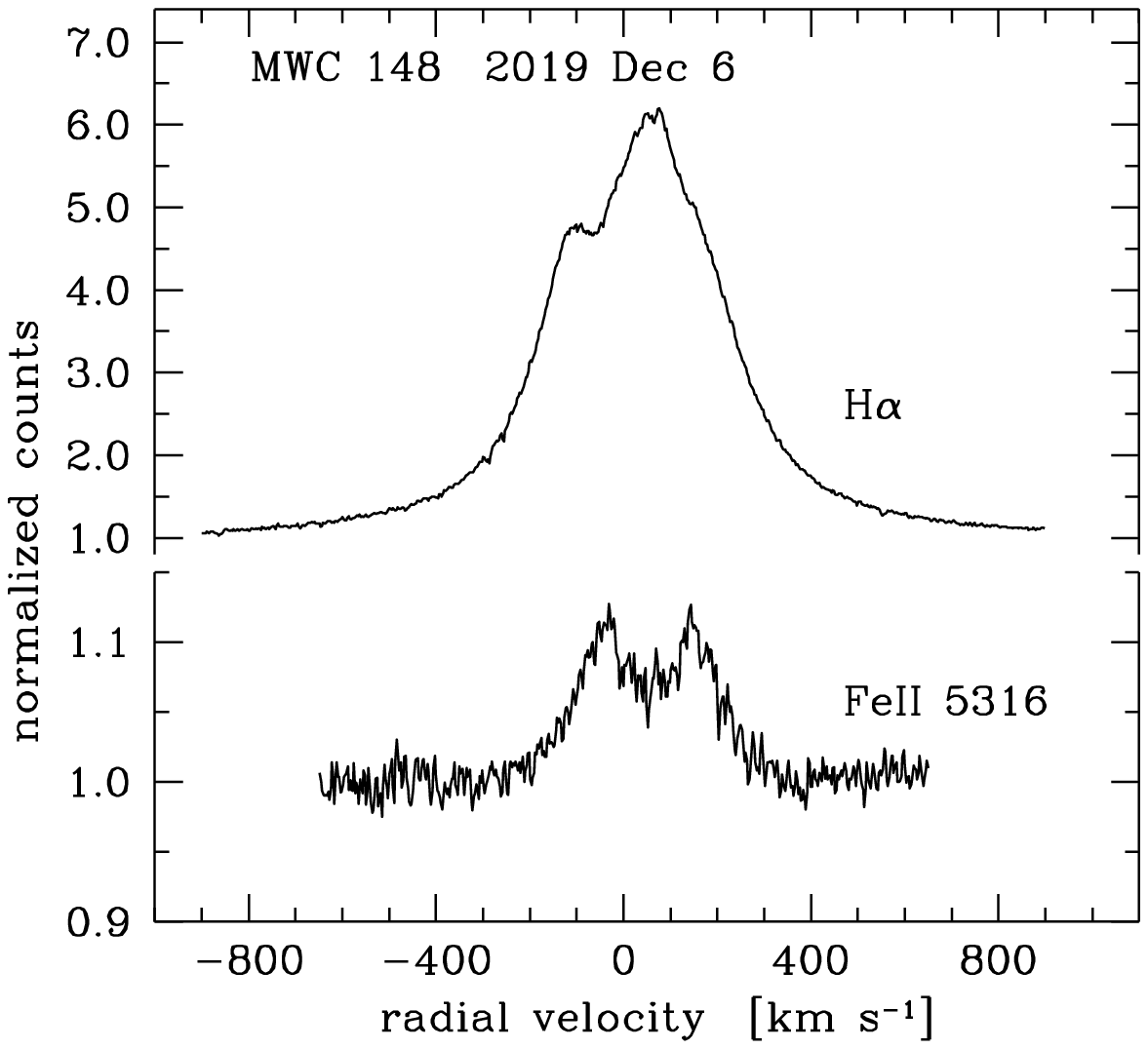}  
   \includegraphics{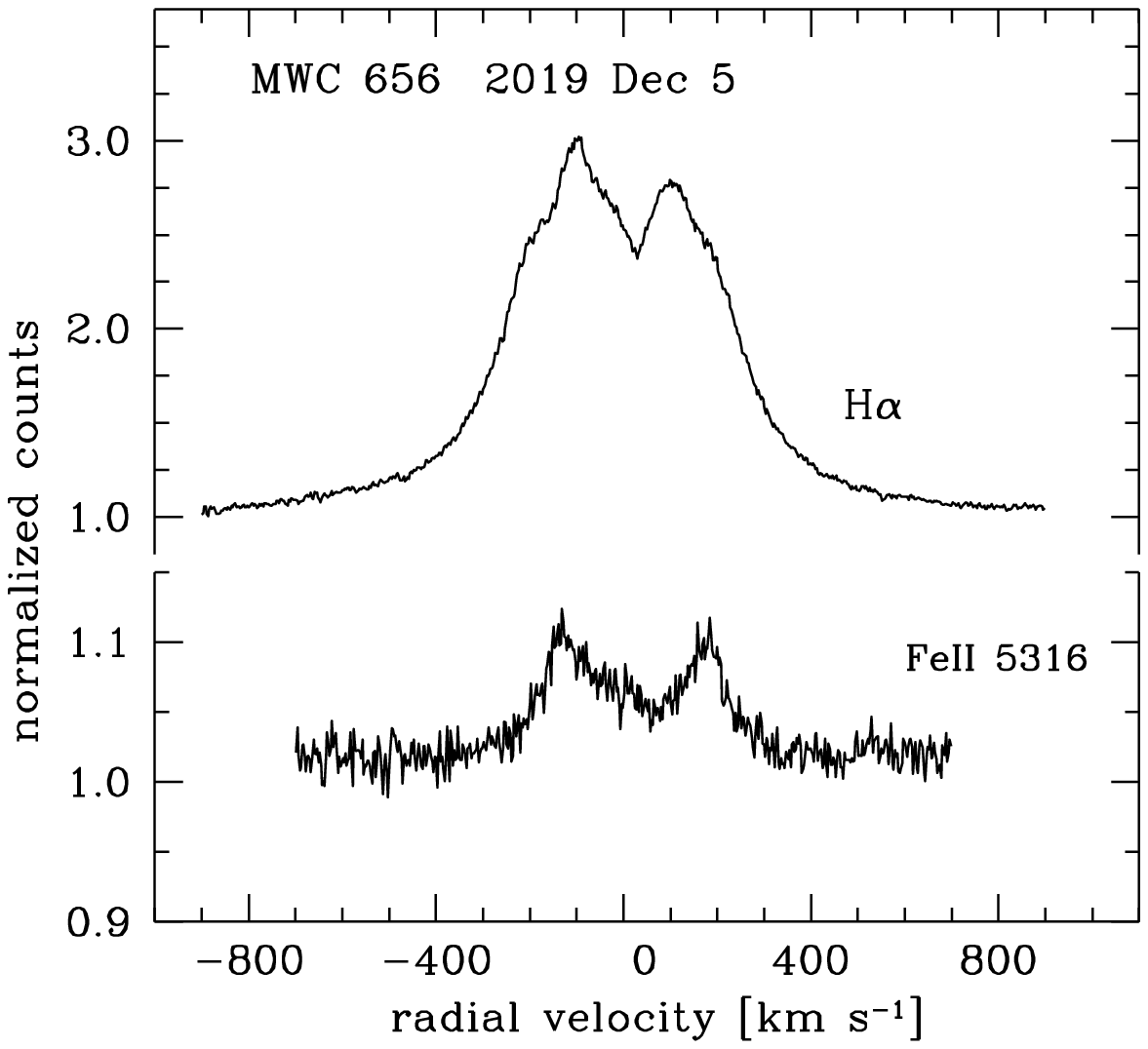}      
   \caption[]{ The emission lines $H\alpha$ and FeII~5316 in the spectra of MWC~148 (left panel) 
    and MWC~656 (right panel). } 
   \label{f.spe}        
 \end{figure*}	     

\section{Observations}


We use  both space photometry and  ground-based spectra.

\subsection{TESS photometric data}
The \textit{Transiting Exoplanet Survey Satellite} (\textit{TESS}, Ricker et al. 2015) is a space-based optical telescope 
launched in 2018 with the primary mission to perform an all-sky survey
to search for transiting exoplanets. In order to fulfill the mission, 
the sky is divided into a number of \textit{sectors}, each of which corresponds 
to the total field of view of all four cameras of the telescope -- $24^\circ\times96^\circ$. 
Each sector is observed for approximately 27 days at a cadence of 2 minutes.
Simple aperture photometry is applied to each of the data files to obtain a barycentered light curve file of the selected object.  
The bandpass of \textit{TESS} is centered on the classical I$_c$ filter, but it is wider and spans from 6000~\AA\ to 10~000~\AA,
in other words the telescope observes the red and the near-infrared emissions of the stars.
The light curves of MWC~656 and MWC~148 were downloaded 
from the Mikulski Archive for Space Telescopes archives\footnote{https://archive.stsci.edu/tess/}.

\subsection{Rozhen spectral data}

The spectral observations were obtained with the  ESPERO spectrograph 
of the 2.0m RCC telescope at Rozhen National Astronomical Observatory located in Rhodope Mountain, Bulgaria. 
ESPERO is a fiber-fed Echelle spectrograph giving a dispersion of 0.06~\AA~px$^{-1}$
and resolving power $\sim 30000$ at 6560~\AA\ (Bonev et al. 2017). 
The spectral  processing and measurements of the spectral lines  
are performed using standard routines provided by IRAF (Tody 1993).
On each spectrum we measure the equivalent width EW(H$\alpha$) and the full width at half maximum (FWHM) 
of the H$\alpha$ emission line, equivalent width EW(5316)
and the full width at zero intensity (FWZI) of the FeII~5316~\AA\  line.
FWZI is the full width of the emission line at the continuum level.
Examples of the emission lines are presented in Fig.~\ref{f.spe}.
They are normalized to the local continuum. 
The typical errors of our measurements 
are $\pm 5$\% in EW(H$\alpha$), 
$\pm 7$~km~s$^{-1}$ in FWHM(H$\alpha$), 
$\pm 10$\% in EW(FeII5316),  
$\pm 40$~km~s$^{-1}$ in FWZI(FeII). 
The log of observations is given in Table~\ref{tab.j}.

Traces of residual wings due to photospheric absorption were not detected and  
the interpolated continuum was taken as the baseline during the measurements of EW and FWHM.
The equivalent width is measured using $splot$ routine in IRAF
by marking two continuum points around the line to be measured. 
The linear continuum is subtracted and the flux is determined by simply summing the pixels 
with partial pixels at the ends.  
The method calculates the area under the profile irrespective of its shape 
(e.g. Mathew \& Subramaniam 2011). 
The FWHM is measured by identifying the points of the emission line profile where the intensity is
equal to one half of the peak intensity, as shown in  Fig.~1 of Glebocki et al. (1986). 
 The horizontal distance between this two points was measured. 
 This measurement also does not depend on the profile shape. 

\section{Analysis methodology}

In this section we give the equations 
connecting the relevant parameters of the primary components on which our estimates are based.
For the Be stars, Hanuschik (1989) gives a relation between projected rotational velocity (${v} \sin i$), 
FWHM(H$\alpha$), and EW(H$\alpha$).  
We use his relation in the form  
 \begin{equation}
   {v} \sin i=0.813 \: (\rm FWHM \: 10^{0.08 \log EW(H\alpha) \ - 0.14} - 70),   
   \label{Han1}
 \end{equation}
where FWHM and  $ v \sin i$ are measured in km~s$^{-1}$; EW(H$\alpha$) is in \AA. 


The Fe II lines are optically thin and their profiles reflect 
the Keplerian rotation in the innermost part of the Be disc (Hanuschik 1996).
The inclination of the Be star is connected with  the full width at zero intensity of the FeII lines
and its radius:  
\begin{equation}
   \frac{\rm FWZI}{2 \sin i }=\left(\frac{\rm GM_1}{(1 + \epsilon) \rm R_1}\right)^{1/2},   \\  
\label{eq.2}
\end{equation}
where G is the gravitational constant, M$_1$ is the mass of the Be star, 
R$_1$ is its radius, $i$ is the inclination of the Be star to the line of sight, $\epsilon$ is a 
dimensionless parameter, $\epsilon \ge 0$. 
Eq.~\ref{eq.2} 
represents the Keplerian motion in the disc and
is a modification of that used in Sect.~6.1 of Casares et al. (2012).
The region where the Fe II lines are produced can be extended down to the very surface of the Be star
or close to it. The parameter $\epsilon$, for which we adopt $ 0 \le \epsilon < 0.1$,
represents how close to the surface of star the emission at FWZI of the FeII lines is formed.

The rotational period of the Be star is also connected with the above parameters: 
\begin{equation}
   {\rm P_{\rm rot}} = \frac { 2 \pi R_1} { {v} \sin i } \sin i .
\label{eq.3}
\end{equation}
The rotational periodicity is  probably due to the interaction between the  magnetic field of the Be
star and its circumstellar disc or the presence of some physical
feature, such as a spot or cloud, co-rotating with the star (Smith, Henry \& Vishniac 2006).
The rotational periods in the Be/$\gamma$-ray binaries are expected to be of order 1~day (Zamanov et al. 2016).

The applied methodology involves the following steps: 
\begin{enumerate}
\item   A periodogram analysis of the TESS data is performed to estimate P$_{\rm rot}$;   \\
\item   The parameter ${v} \sin i $ is estimated using Eq.~\ref{Han1} and the data in Table~\ref{tab.sp}; \\
\item   A mass value for the Be star is adopted according to its  spectral type; \\
\item   Using Eq.~\ref{eq.2} and  Eq.~\ref{eq.3}, the values of R$_1$ and $i$ are calculated for
           the primaries of MWC~148 and MWC~656.  
\end{enumerate}
The period-search methods applied to the TESS photometry 
were the phase dispersion minimization, PDM (Stellingwerf 1978) 
and the CLEAN algorithm (Roberts, Lehar \& Dreher 1987).

\section{ MWC 148} 
\begin{figure*}    
   \vspace{5.3cm}     
   \includegraphics{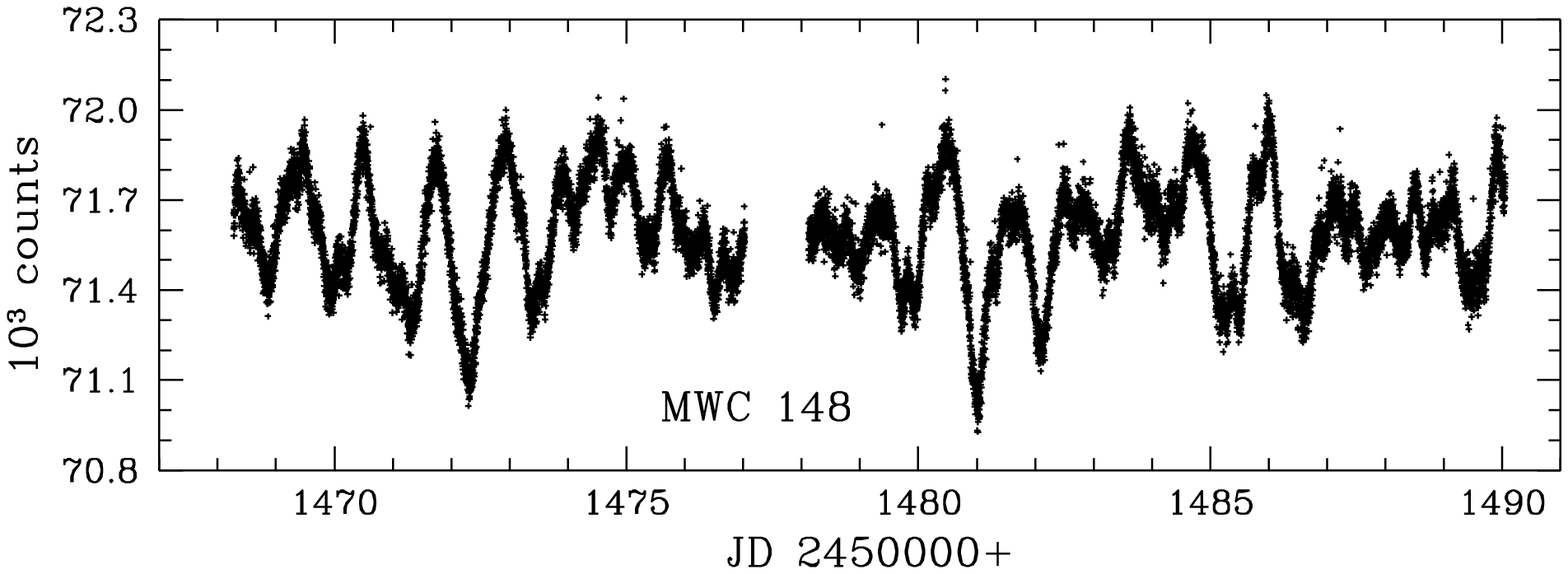} 
   \caption[]{TESS light curve of MWC 148.  } 
   \label{f.MWC148a} 
   \vspace{5.0cm}      
   \includegraphics{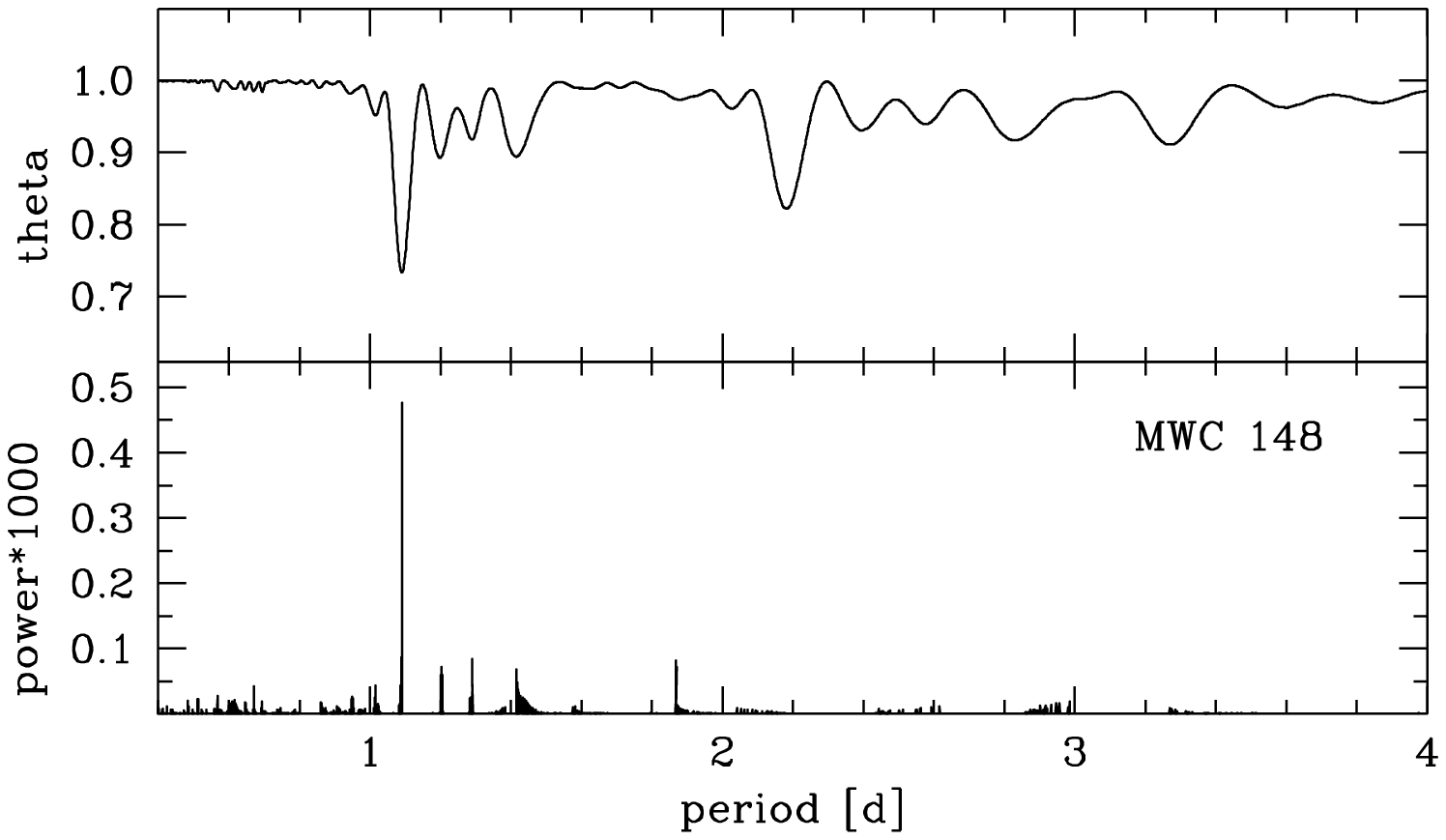}  
   \includegraphics{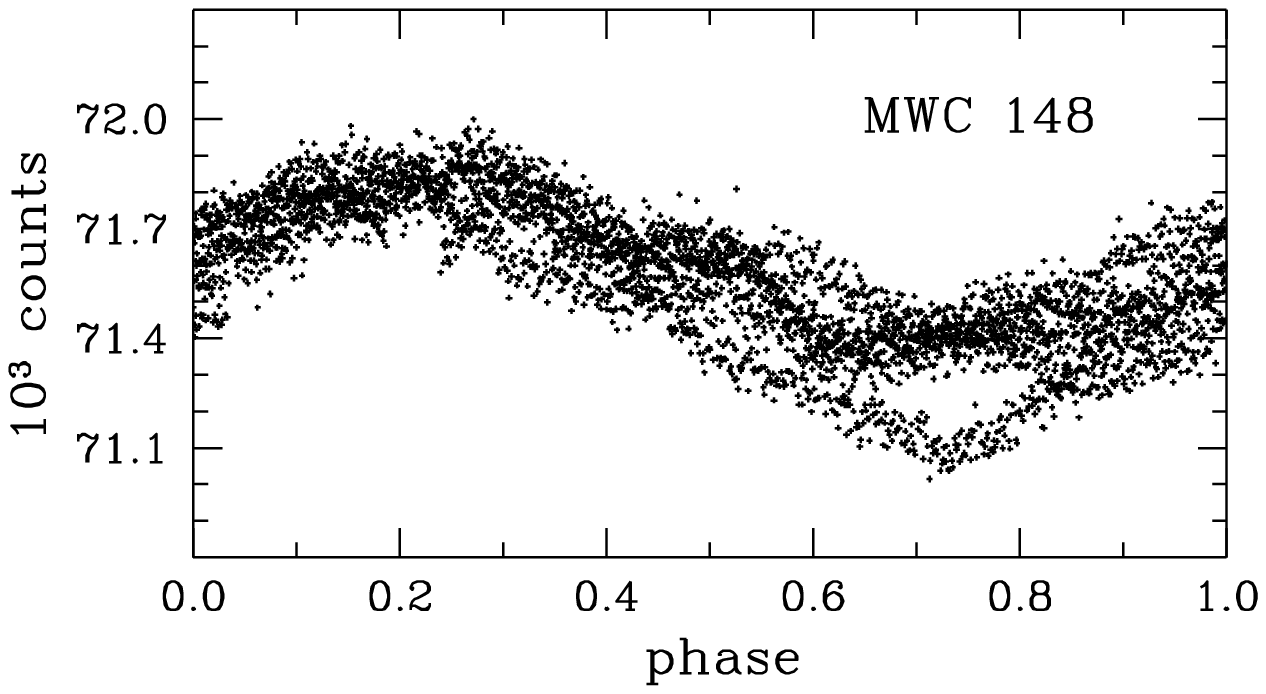}      
   \caption[]{Left panel: periodogram analysis for MWC~148 --  theta  and power versus period. 
    The right panel represents the light curve during 1468 - 1474
    folded with a 1.156 d  period. } 
   \label{f.MWC148b}        
   \vspace{5.6cm}     
   \includegraphics{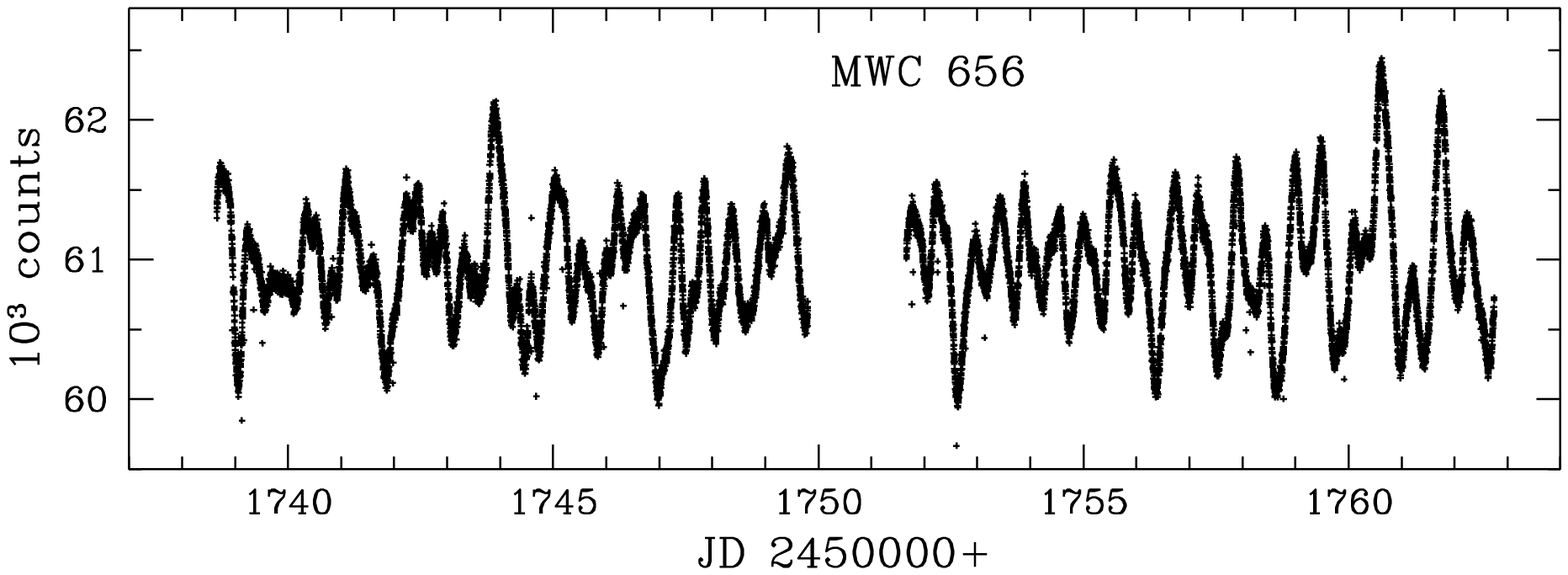} 
   \caption[]{TESS light curve of MWC~656. } 
   \label{f.MWC656a} 
   \vspace{5.0cm}      
   \includegraphics{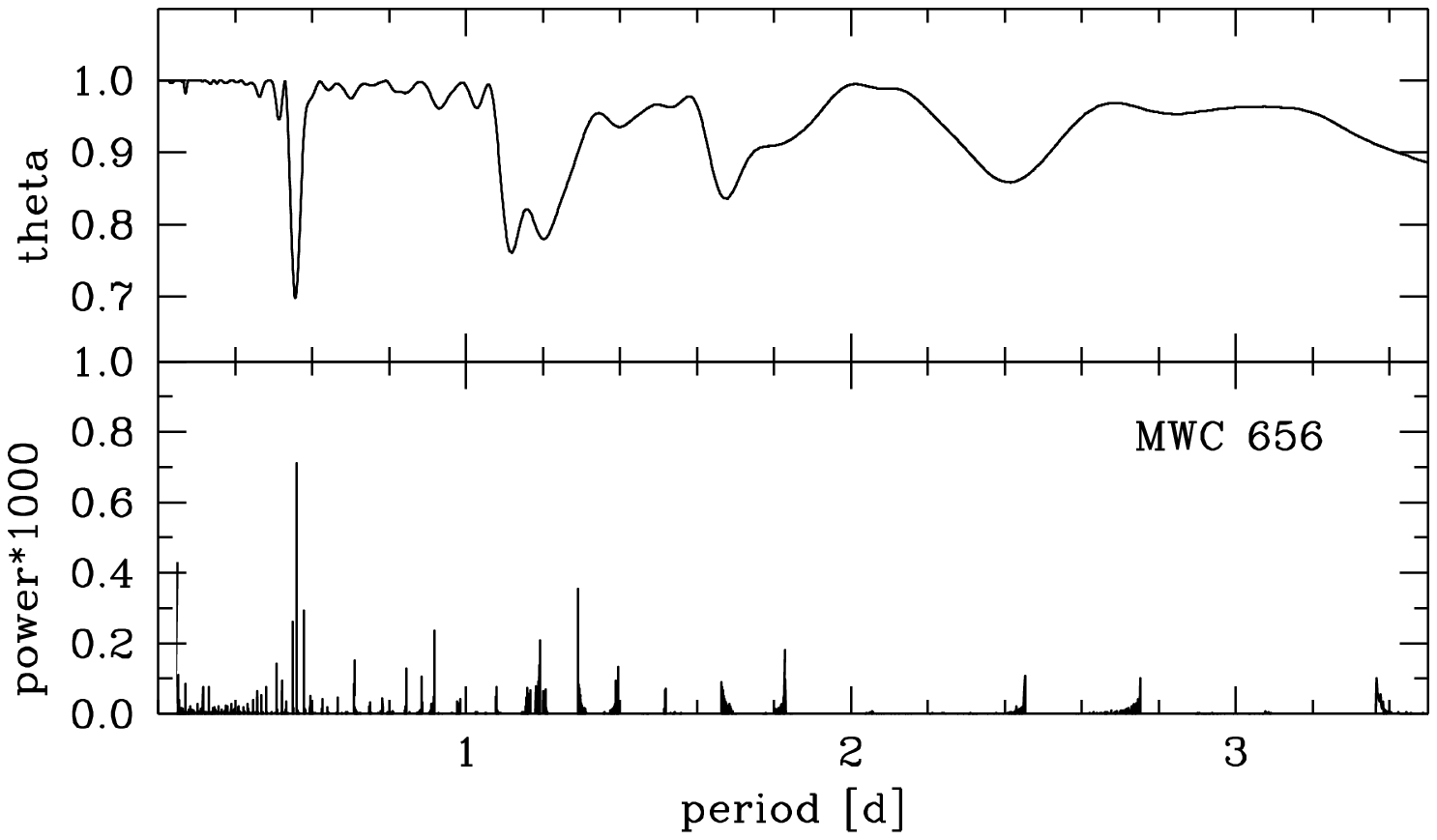}  
   \includegraphics{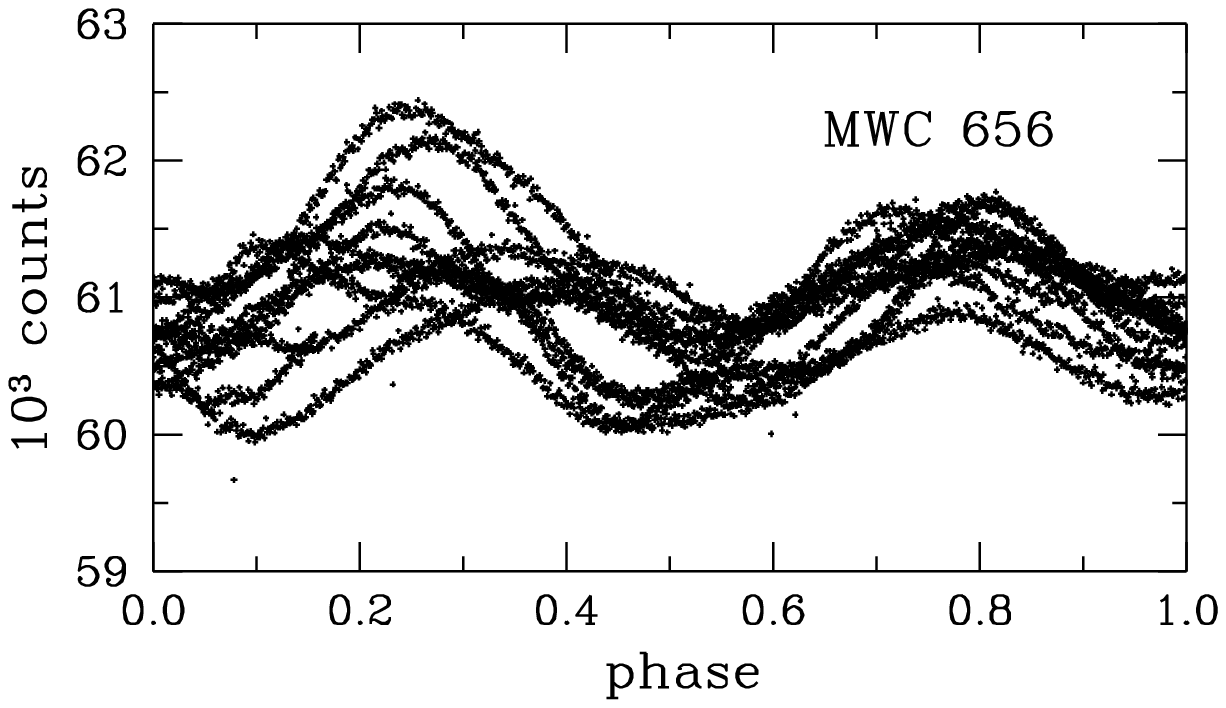}      
   \caption[]{Left panel: periodogram analysis for MWC~148. 
    The most significant period is 0.559 days.  The right panel represents   
    the light curve during 1751 - 1763
    folded with a 1.117 d  period, which is probably the rotational period of the B star. 
    } 
   \label{f.MWC656b}         
 \end{figure*}

TESS photometry  for MWC 148, in the interval JD~2451468.2 -- JD~2451490.0, 
can be accessed under Input Catalogue ID 234929785 
and is plotted in Fig.~\ref{f.MWC148a}. 
The periodogram analysis is presented in the left panel of Fig.~\ref{f.MWC148b}, where the PDM statistic 
(theta) and the CLEAN component amplitude are plotted.
Over the entire data set, the analysis yields P$_{\rm rot} = 1.0908 \pm 0.0002$~d.      
The periodogram analysis of the data in the $({\rm JD} - 2450000)$ time interval  
1468 -- 1490  gives a clear  period of 1.09 $\pm 0.03$ days, 
in the interval 1468 -- 1477  a clear period of 1.131 $\pm 0.025$ days,
and in the interval 1478 -- 1490  a clear  period of 1.075 $\pm 0.025$ days. 
A visual inspection of the data gives us the possibility to select
the parts of the light curve where this periodicity is most clearly visible. 
Using  days 1468 -- 1474, we find 1.156 d,  and for days 1479 -- 1487 we find 1.101 d. 
The light curve 
for days 1468 -- 1474 is plotted
in the right panel of Fig.~\ref{f.MWC148b}.  
We consider that the rotational period of the Be star in MWC~148 is 
in the range $ 1.09 \le \rm P_{rot} < 1.16$~d. 

Using Eq.~\ref{Han1}  and the  values given in Table~\ref{tab.sp}, we find   $ {v} \sin i =272 \pm 5$~km~s$^{-1}$. 
Casares et al. (2012) give a spectral type  B0Vpe. 
A B0V star is expected to have on average  mass 
M$_1 = 15.0 \pm 0.5$ ~M$_\odot$ (Hohle et al. 2010). 
For MWC~148, Aragona et al. (2010) derived
M$_1 = 13.2 - 19.0$~M$_\odot$ from spectral model fits, which is in agreement with 
the adopted M$_1 = 15.0$~M$_\odot$.
Adopting  FWZI($\lambda 5316$) $\approx 694$~km~s$^{-1}$,  
and  ${\rm v} \sin i = 272$~km~s$^{-1}$, 
we find   $i = 40^\circ \pm 2^\circ$
and  R$_1= 9.2 \pm 0.5$~R$_\odot$. 

It is worth noting for comparison that 
(i) a B0V star is expected to have R$_1 = 7.2$ $R_{\odot}$ (Straizys \& Kuriliene 1981); 
(ii) Casares et al. (2012) give  FWZI($\lambda 5018$) $\sim 1300$~km~s$^{-1}$
and $ {v} \sin i = 373$~km~s$^{-1}$;
(iii) $ {v} \sin i = 430$~km~s$^{-1}$ (Guti{\'e}rrez-Soto et al. 2007) 
and $ {v} \sin i = 500$~km~s$^{-1}$ (Aragona et al. 2010) are reported for this object. 

\section{ MWC 656} 

TESS photometry for
MWC 656 (HD 215227), in the interval JD~2451738 -- 2451763, can be accessed under
Input Catalogue ID 153880067 (Fig.~\ref{f.MWC656a}).
The data gap is due to the telescope being repointed to
transfer data to Earth at this time.   
The periodogram analysis is shown in the left panel of Fig.~\ref{f.MWC656b} with the same kind 
of plots as for the previous source.
Over the entire data set, the analysis yields P$_{\rm rot} = 0.5586 \pm 0.0002$~d.  
In contrast, we do not find a clear period   
in the $({\rm JD} - 2450000)$ interval  1738 -- 1751 days.
However in the interval  1751.6  -- 1762.7 days,  a clear  period of 0.557 $\pm 0.025$ days is detected.
Looking at the interval 1760-1763 days, we see that there is a repetition of strong and 
weak maxima, which probably means that the rotational period is 
doubled P$_{\rm rot}=1.117$~d.

For the primary, Williams et al. (2010)
estimated T$_{\rm eff} = 19 000 \pm 3000$ K, $\log g = 3.7 \pm 0.2$, 
M$_1 = 7.7 \pm 2.0$~M$_\odot$ , and R$_1 = 6.6 \pm 1.9$~R$_\odot$.   
Casares et al. (2014)  found that the mass donor is a giant (B1.5-2 III) and gave a
mass range 10--16~M$_\odot$.
From the results of Hohle et al. (2010), such a star is expected to have $8.0 < \rm M_1 < 10.0$~M$_\odot$.


Using the relation between  ${v} \sin i$, 
FWHM($H\alpha$), and EW(H$\alpha$) (Eq.~\ref{Han1}),  
and our values (see Table~\ref{tab.sp}), we find   $ {v} \sin i =313 \pm 3$~km~s$^{-1}$,
which is similar to the values $330 \pm 30$~km~s$^{-1}$ (Casares et al. 2014),
$262 \pm 26$~km~s$^{-1}$ (Yudin 2001), and  $330 \pm50$~km~s$^{-1}$  (Williams et al. 2010). 
 
We  measure  FWZI(FeII 5316) = 12.5-12.9 \AA $= 709 \pm 12$~km~s$^{-1}$,
which is below the value FWZI(FeII 5018)$ \sim 1000$~km~s$^{-1}$ (Casares et al. 2012). 
 
Adopting  M$_1 = 10.0 \pm 0.5$~M$_\odot$, $ {v} \sin i =313$~km~s$^{-1}$,
and $FWZI (FeII) = 709$~km~s$^{-1}$,  the system of  Eqs.~\ref{eq.2} and  \ref{eq.3} can be solved. As a result, 
we find  R$_1=8.8 \pm 0.5 $~R$_\odot$ and  $i =52^\circ \pm 2^\circ$.    
The obtained value of  R$_1$ agrees with the estimates of the average radius for a
B1.5-2 III star ($8.3 - 8.8$ R$_\odot$, Straizys \& Kuriliene 1981).

For MWC 656, from the radial velocity measurements, 
Casares et al. (2014) obtained $M_1 \sin ^3 i_{\rm orb} = 5.83 \pm  0.70$. 
For  M$_1 = 10$~M$_\odot$, this gives $53^\circ < i_{\rm orb}< 60^\circ$. 
It appears that the orbital plane and the equatorial 
plane of the Be star are practically coplanar within $\pm 4^\circ$.

%

\section{Discussion}

In  Be/X-ray binaries the primary is a rapidly  rotating Be star with mass $\sim 10$~M$_\odot$ 
and the secondary is a neutron star or a black hole. The secondary mass is expected to be 
$\sim 2$~M$_\odot$ for a neutron star, and  $\sim 5$~M$_\odot$ in the case of black hole. 
The orbital periods are in the range  10 -- 400 d  (Reig 2011). 
The Be/$\gamma$-ray binaries are a subgroup of the Be/X-ray binaries 
and should have similar binary parameters. 
They are a product of the evolution of a binary
containing two moderately massive stars, which undergoes mass transfer from
the originally more massive star towards its companion (Pols et al. 1991; Negueruela 2007). 
The eccentricities in these systems are  caused by a kick to
the compact object  during the supernova explosion that formed it (e.g. Martin et al. 2009). 


HESS J0632+057 (MWC~148) produces non-thermal radio, X-ray, GeV and 
very high-energy gamma-ray emission.  The non-thermal emission 
is modulated with the 315 days orbital period and has a a peculiar light curve containing two peaks, separated by a dip --
a sharp peak with a short decay before the apastron passage and a broad (several tens of days) 
secondary peak after the apastron passage in X-rays and TeV (Aliu et al. 2014, Archer et al. 2020). 
The non-thermal activity before and around apastron can be linked to
(i) the accumulation of non-thermal particles 
in the vicinity of the binary, and the sudden drop of the emission before apastron 
is produced by the disruption of the two-wind interaction structure, allowing these particles to escape efficiently
(Bosch-Ramon et al. 2017) or 
(ii) an accumulation of hot shocked plasma in the inner spiral arm, later released when the spiral arm is disrupted 
in the periastron-apastron direction (Barkov \& Bosch-Ramon 2018). 

For MWC~148,  two solutions for the orbit indicate that it 
is highly eccentric: $e = 0.83 \pm 0.08$ (Casares et al. 2012)
and $e = 0.64 \pm 0.29$  (Moritani et al. 2018). 
In our previous paper (Zamanov et al. 2017), 
we assumed the value of $i$, on the basis of the strong resemblance
of the optical emission lines between MWC 148 and the bright Be star $\gamma$~Cas (Zamanov, Stoyanov \& Mart{\'\i} 2016), 
for which the inclination is $i = 43^\circ \pm 3^\circ$ (Poeckert \& Marlborough 1978; Clarke 1990). 
Our result here for MWC~148 ($i = 40^\circ  \pm 2^\circ $) confirms the assumption 
that one of the reasons for this similarity is the inclination.

MWC 656 is faint in X-rays and it reaches the faintest X-ray luminosities ever detected in stellar-mass black holes
(Rib\'o et al. 2017). It may not continuously emit in $\gamma$-rays  (Alexander \& McSwain 2016). 
For this binary, Casares et al. (2014) found that the mass of the black hole 
is in the range 3.8 -- 6.9~\msun.
The orbital eccentricity is $e$ = 0.10 $\pm$ 0.04.
It is suggested that the warp and precession observed
in Be star discs may be caused by a misalignment between
the spin axis of the Be star and the orbit of the binary companion (Martin et al. 2011).
Our results indicate that there is no misalignment in MWC~656, and consequently 
warping and precession should not be observed. 

The velocity kick from the supernova has two effects -- it renders the orbit eccentric 
and it misaligns the orbit with respect to the spin axis of the Be star.  
In systems that experience
low velocity kicks, the misalignments tend to be small  (Martin et al. 2011). 
The low orbital eccentricity and the alignment between the spin axis of the primary  
and the axis of the binary orbit, 
indicates that the compact object in MWC~656 was born with low kick velocity.


\section{Conclusions:}  
We evaluated  some parameters for the mass donor stars in the Be/$\gamma$-ray binaries MWC~148 and MWC~656. 
For  MWC~148, we estimate
  $ {v} \sin i =272 \pm 5$~km~s$^{-1}$, P$_{\rm rot} = 1.10 \pm 0.03$~d, R$_1= 9.2 \pm 0.5$~R$_\odot$, and $i = 40^\circ \pm 2^\circ$. 
For  MWC~656, we obtain 
$ {v} \sin i =313 \pm 3$~km~s$^{-1}$,
P$_{\rm rot} = 1.12 \pm 0.03$~d, R$_1= 8.8 \pm 0.5$~R$_\odot$, and $i = 52^\circ \pm 3^\circ$.
These parameters should be useful for future accurate modeling of these systems. 


\acknowledgements
 
 This work is supported by Bulgarian National Science Fund -- project K$\Pi$-06-H28/2
 "Binary stars with compact object". 
 JM acknowledges support  by grant PID2019-105510GB-C32 / AEI / 10.13039/501100011033 from 
 the Agencia Estatal de Investigaci\'on of the Spanish Ministerio de Ciencia, Innovaci\'on y Universidades, 
 and by Consejer\'{\i}a de Econom\'{\i}a, Innovaci\'on, Ciencia y Empleo of Junta de Andaluc\'{\i}a 
 as research group FQM- 322, as well as FEDER funds.
 YMN acknowledges grant DCM~577/17.08.2018 from National Research Programme "Young
 scientists and postdoctoral students" of  the Bulgarian Ministry of Education and Science. 
We are grateful to Mike Bode (Liverpool John Moores University) for  critical reading of the manuscript
and the anonymous referee for useful comments.

%

%

\begin{thebibliography}{}
\bibitem[Abdo et al.(2009)]{2009Sci...325..840A} Abdo, A.~A., Ackermann, M., Ajello, M., et al.\ 2009, Science, 325, 840
\bibitem[Aharonian et al.(2005)]{2005A&A...442....1A} Aharonian, F., Akhperjanian, A.~G., Aye, K.-M., et al.\ 2005a,\aap, 442, 1
\bibitem[Aharonian et al.(2005)]{2005Sci...309..746A} Aharonian, F., Akhperjanian, A.~G., Aye, K.-M., et al.\ 2005b, Science, 309, 746
\bibitem[Aharonian et al.(2007)]{2007A&A...469L...1A} Aharonian, F.~A., Akhperjanian, A.~G., Bazer-Bachi, A.~R., et al.\ 2007, \aap, 469, L1
\bibitem[Albert et al.(2006)]{2006Sci...312.1771A} Albert, J., Aliu, E., Anderhub, H., et al.\ 2006, Science, 312, 1771
\bibitem[Aleksi{\'c} et al.(2015)]{2015A&A...576A..36A} Aleksi{\'c}, J., Ansoldi, S., Antonelli, L.~A., et al.\ 2015, \aap, 576, A36
\bibitem[Alexander \& McSwain(2016)]{2016ASPC..506..243A} Alexander, M.~J. \& McSwain, M.~V.\ 2016, ASP Conf. Ser., 506, 243
\bibitem[Aliu et al.(2014)]{2014ApJ...780..168A} Aliu, E., Archambault, S., Aune, T., et al.\ 2014, \apj, 780, 168
\bibitem[Aragona et al.(2010)]{2010ApJ...724..306A} Aragona, C., McSwain, M.~V., \& De Becker, M.\ 2010, \apj, 724, 306
\bibitem[Archer et al.(2020)]{2020ApJ...888..115A}  Archer, A., Benbow, W.,  Bird, R. et al.\ 2020, \apj, 888, 115
\bibitem[Barkov \& Bosch-Ramon(2018)]{2018MNRAS.479.1320B} Barkov, M.~V. \& Bosch-Ramon, V.\ 2018, \mnras, 479, 1320
\bibitem[Bonev et al.(2017)]{2017BlgAJ..26...67B} Bonev, T., Markov, H., Tomov, T., et al.\ 2017, Bulgarian Astronomical Journal, 26, 67
\bibitem[Bosch-Ramon et al.(2017)]{2017MNRAS.471L.150B} Bosch-Ramon, V., Barkov, M.~V., Mignone, A., et al.\ 2017, \mnras, 471, L150
\bibitem[Casares et al.(2014)]{2014Natur.505..378C} Casares, J., Negueruela, I., Rib{\'o}, M., et al.\ 2014, \nat, 505, 378
\bibitem[Casares et al.(2012)]{2012MNRAS.421.1103C} Casares, J., Rib{\'o}, M., Ribas, I., et al.\ 2012, \mnras, 421, 1103
\bibitem[Chernyakova \& Malyshev(2020)]{2020arXiv200603615C} Chernyakova, M. \& Malyshev, D.\ 2020,  Proceeding of Science, 362, 045 
\bibitem[Clarke(1990)]{1990A&A...227..151C} Clarke, D.\ 1990, \aap, 227, 151
\bibitem[Corbet et al.(2011)]{2011ATel.3221....1C} Corbet, R.~H.~D., Cheung, C.~C., Kerr, M., et al.\ 2011, The Astronomer's Telegram, 3221
\bibitem[Corbet et al.(2016)]{2016ApJ...829..105C} Corbet, R.~H.~D., Chomiuk, L., Coe, M.~J., et al.\ 2016, \apj, 829, 105
\bibitem[Dubus(2006)]{2006A&A...456..801D} Dubus, G.\ 2006, \aap, 456, 801
\bibitem[Dubus(2013)]{2013A&ARv..21...64D} Dubus, G.\ 2013, \aapr, 21, 64
\bibitem[Glebocki et al.(1986)]{1986A&A...158..392G} Glebocki R., Sikorski J., Bielicz E., Krogulec M., 1986, \aap 158, 392
\bibitem[Guti{\'e}rrez-Soto et al.(2007)]{2007A&A...476..927G} Guti{\'e}rrez-Soto, J., Fabregat, J., Suso, J., et al.\ 2007, \aap, 476, 927   
\bibitem[Hanuschik(1989)]{1989Ap&SS.161...61H} Hanuschik, R.~W.\ 1989, Ap\&SS, 161, 61 
\bibitem[Hanuschik(1996)]{1996A&A...308..170H} Hanuschik, R.~W.\ 1996, \aap, 308, 170
\bibitem[Hohle et al.(2010)]{2010AN....331..349H} Hohle, M.~M., Neuh{\"a}user, R., \& Schutz, B.~F.\ 2010, \an, 331, 349
\bibitem[Johnston et al.(1992)]{1992ApJ...387L..37J} Johnston, S., Manchester, R.~N., Lyne, A.~G., et al.\ 1992, \apjl, 387, L37
\bibitem[Li et al.(2017)]{2017ApJ...846..169L} Li J., Torres D.~F., Cheng K.-S., et al. \ 2017, \apj, 846, 169
\bibitem[Lucarelli et al.(2010)]{2010ATel.2761....1L} Lucarelli, F., Verrecchia, F., Striani, E., et al.\ 2010, The Astronomer's Telegram, 2761
\bibitem[Lyne et al.(2015)]{2015MNRAS.451..581L} Lyne, A.~G., Stappers, B.~W., Keith, M.~J., et al.\ 2015, \mnras, 451, 581
\bibitem[Martin et al.(2009)]{2009MNRAS.397.1563M} Martin, R.~G., Tout, C.~A., \& Pringle, J.~E.\ 2009, \mnras, 397, 1563 
\bibitem[Martin et al.(2011)]{2011MNRAS.416.2827M} Martin, R.~G., Pringle, J.~E., Tout, C.~A., et al.\ 2011, \mnras, 416, 2827 
\bibitem[Massi \& Jaron(2013)]{2013A&A...554A.105M} Massi, M. \& Jaron, F.\ 2013, \aap, 554, A105
\bibitem[Mathew \& Subramaniam(2011)]{2011BASI...39..517M} Mathew, B. \& Subramaniam, A.\ 2011, Bulletin of the Astronomical Society of India, 39, 517
\bibitem[Mirabel(2012)]{2012Sci...335..175M} Mirabel, I.~F.\ 2012, Science, 335, 175
\bibitem[Moritani et al.(2018)]{2018PASJ...70...61M} Moritani, Y., Kawano, T., Chimasu, S., et al.\ 2018, \pasj, 70, 61
\bibitem[Negueruela(2007)]{2007ASPC..367..477N} Negueruela, I.\ 2007, ASP Conf. Ser., 367, 477
\bibitem[Paredes \& Bordas(2019)]{2019arXiv190209898P} Paredes, J.~M. \& Bordas, P.\ 2019, Rendiconti Lincei. Scienze Fisiche e Naturali,  30, 107
\bibitem[Poeckert \& Marlborough(1978)]{1978ApJ...220..940P} Poeckert, R. \& Marlborough, J.~M.\ 1978, \apj, 220, 940
\bibitem[Pols et al.(1991)]{1991A&A...241..419P} Pols, O.~R., Cote, J., Waters, L.~B.~F.~M., et al.\ 1991, \aap, 241, 419
\bibitem[Reig(2011)]{2011Ap&SS.332....1R} Reig, P.\ 2011, \apss, 332, 1
\bibitem[Rib{\'o} et al.(2017)]{2017ApJ...835L..33R} Rib{\'o}, M., Munar-Adrover, P., Paredes, J.~M., et al.\ 2017, \apjl, 835, L33
\bibitem[Ricker et al.(2015)]{2015JATIS...1a4003R} Ricker, G. R., Winn, J. N., Vanderspek, R., et al. 2015, Journal of Astronomical  Telescopes, Instruments, and Systems, 1, 014003
\bibitem[Roberts et al.(1987)]{1987AJ.....93..968R} Roberts, D.~H., Lehar, J., \& Dreher, J.~W.\ 1987, \aj, 93, 968 
\bibitem[Romero et al.(2007)]{2007A&A...474...15R} Romero, G.~E., Okazaki, A.~T., Orellana, M., et al.\ 2007, \aap, 474, 15
\bibitem[Smith et al.(2006)]{2006ApJ...647.1375S} Smith, M.~A., Henry, G.~W., \& Vishniac, E.\ 2006, \apj, 647, 1375
\bibitem[Stellingwerf(1978)]{1978ApJ...224..953S} Stellingwerf, R.~F., 1978, \apj, 224, 953
\bibitem[Straizys \& Kuriliene(1981)]{1981Ap&SS..80..353S} Straizys, V. \& Kuriliene, G.\ 1981, \apss, 80, 353
\bibitem[Tody(1993)]{1993ASPC...52..173T} Tody, D.\ 1993, ASP Conf. Ser.,  52, 173
\bibitem[Wang \& Robertson(1985)]{1985A&A...151..361W} Wang, Y.-M. \& Robertson, J.~A.\ 1985, \aap, 151, 361
\bibitem[Williams et al.(2010)]{2010ApJ...723L..93W} Williams, S.~J., Gies, D.~R., Matson, R.~A., et al.\ 2010, \apjl, 723, L93
\bibitem[Yudin(2001)]{2001A&A...368..912Y} Yudin, R.~V.\ 2001, \aap, 368, 912
\bibitem[Zamanov  et al.(2016)]{2016BlgAJ..24...40Z} Zamanov, R., Stoyanov, K., \& Mart{\'\i}, J., 2016, Bulgarian Astronomical Journal, 24, 40
\bibitem[Zamanov et al.(2016)]{2016A&A...593A..97Z} Zamanov, R.~K., Stoyanov, K.~A., Mart{\'\i}, J., et al.\ 2016, \aap, 593, A97
\bibitem[Zamanov  et al.(2017)]{2017BlgAJ..27...57Z} Zamanov, R., Mart{\'\i}, J., \& Garc{\'\i}a-Hern{\'a}ndez, M.~T., 2017, Bulgarian Astronomical Journal, 27, 57 
\end{thebibliography}
%

\end{document}